\documentclass[acmsmall,screen]{acmart}
\AtBeginDocument{%
  }

\usepackage{multirow}
\usepackage{xspace}
\usepackage{xcolor}
\usepackage{xcolor,colortbl}
\usepackage{tabularx}
\usepackage{caption}

\definecolor{amber}{rgb}{1.0, 0.49, 0.0}

\usepackage{enumitem}
\usepackage{framed}
\definecolor{formalshade}{rgb}{0.95, 0.95, 1}
\definecolor{mygray}{gray}{0.4}
\definecolor{lightgray}{gray}{0.93}

\setcopyright{none}
\renewcommand\footnotetextcopyrightpermission[1]{} 
\setcopyright{acmlicensed}
\copyrightyear{2018}
\acmYear{2018}
\acmDOI{XXXXXXX.XXXXXXX}

\acmJournal{JACM}
\acmVolume{37}
\acmNumber{4}
\acmArticle{111}
\acmMonth{8}

\begin{document}

\title[How Smart Is Your GUI Agent?]{How Smart Is Your GUI Agent? A Framework for the Future of Software Interaction}


\author{Sidong Feng}
\affiliation{%
  \institution{The Chinese University of Hong Kong}
  \city{Shenzhen}
  \country{China}}
\email{sidong.feng@monash.edu}

\author{Chunyang Chen}
\affiliation{%
  \institution{Technical University of Munich}
  \city{Heilbronn}
  \country{Germany}
}
\email{chun-yang.chen@tum.de}

\renewcommand{\shortauthors}{Sidong Feng \& Chunyang Chen}

\begin{abstract}
GUI agents are rapidly becoming a new interaction to software, allowing people to navigate web, desktop and mobile rather than execute them click by click. Yet ``agent'' is described with radically different degrees of autonomy, obscuring capability, responsibility and risk. We call for conceptual clarity through GUI Agent Autonomy Levels (GAL), a six-level framework that makes autonomy explicit and helps benchmark progress toward trustworthy software interaction.
\end{abstract}

\settopmatter{printacmref=false}



\begin{teaserfigure}
\centering
\includegraphics[width=0.99\textwidth]{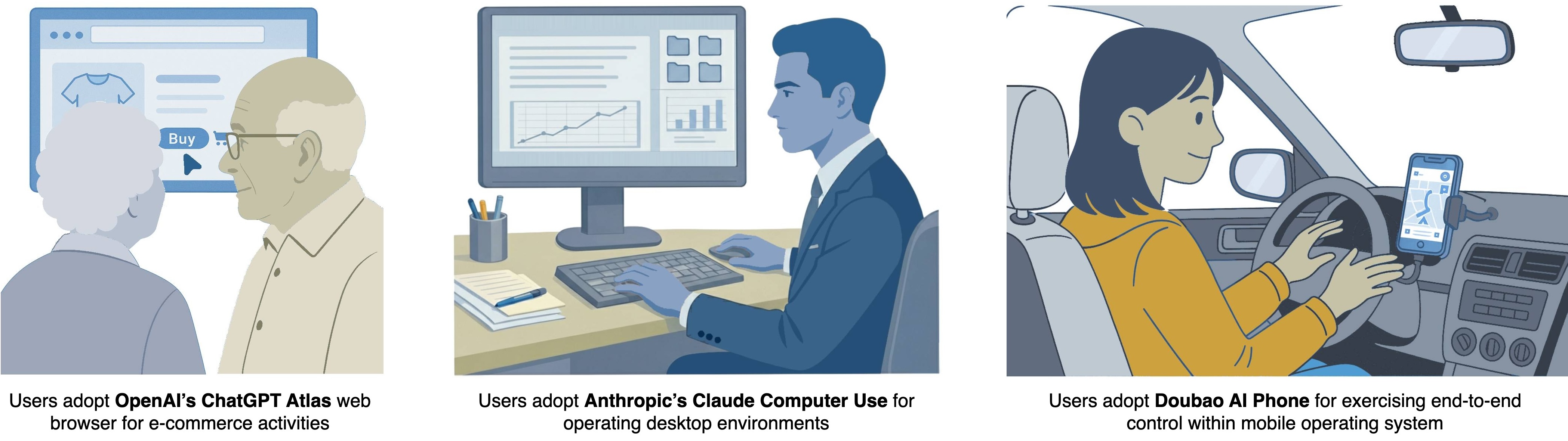}
  \caption{Users rely on a GUI agent in scenarios for the web (left), desktop (middle), and mobile (right).}
  \label{fig:scenario}
\end{teaserfigure}

\maketitle


\section{Introduction}

The Graphical User Interfaces (GUIs) have become a global front door for modern software, mediating nearly every digital task, from writing documents and managing files to navigating enterprise systems and configuring applications. 
Their visual, flexible nature has made them indispensable for human users, but these same qualities make them difficult for machines to interact with. 
Unlike APIs, GUIs were never designed for autonomous interaction: their layouts shift across devices, elements move or re-render dynamically, and meaningful actions often depend on visual interpretation rather than stable programmatic execution.
As software ecosystems grow more complex and fragmented, users increasingly face repetitive, multi-step workflows that remain manual despite major advances in digital automation.
This gap has motivated the emergence of GUI agents, that is, an entity designed to understand user objectives, perceive interface elements, and act directly within GUIs on behalf of users.

Early generations of GUI agents emerged as extensions of software automation efforts, providing foundational capabilities for interacting with interfaces programmatically, like Selenium~\cite{selenium} and AppleScript~\cite{applescript_guide}, relied on predefined workflows and rigid selectors. Although effective for scripted testing or routine task replication, they offered little adaptability to dynamic layouts or ambiguous user goals. 
The recent GUI agents, powered by large multimodal models, can intelligently interpret interface layouts and autonomously execute multi-step actions across diverse software environments. Research prototypes such as WebAgent~\cite{gurreal} and UI-TARS~\cite{qin2025ui} demonstrate that GUI agents can jointly ground language, vision, and action. Complementing these advances, open-source benchmarks like OSWorld~\cite{xie2024osworld} and AndroidWorld~\cite{rawlesandroidworld} provide extensive evaluation environments that enable systematic measurement of agent capabilities and track improvements in complex GUIs.

Meanwhile, industry has begun to deploy GUI agents at scale, delivering tangible benefits to everyday digital tasks and real-life workflow, beyond research settings. Emerging AI browsers, most notably OpenAI’s ChatGPT Atlas~\cite{altas} and Perplexity Comet~\cite{comet}, demonstrate how agents can autonomously navigate the unprecedented scale, semantic heterogeneity, and dynamic web content to assist with user tasks, e.g., e-commerce product comparison and checkout. Anthropic’s Claude Computer Use~\cite{claude_computer_use} further illustrates the maturation of these agents by operating full desktop environments, where it can compose documents, manipulate files, analyze data, and coordinate actions across applications. Beyond desktop and browser contexts, AI-integrated devices such as the Doubao AI Phone embed agentic capabilities directly into the operating system, enabling GUI agents to exercise end-to-end control within mobile interfaces, manage apps, and support on-device personal workflows.

The expanding set of research prototypes, open-source benchmarks, and industrial-grade GUI agents has accelerated widespread exploration of this emerging field, accompanied by a number of surveys, technical reports, and blogs attempting to map the fast-moving landscape. Yet, the field still lacks a shared vocabulary for describing how autonomous a GUI agent truly is. 
Inspired by the autonomy levels like SAE levels~\cite{sae}, this paper introduces the \textit{\textbf{G}UI \textbf{A}gent Autonomy \textbf{L}evels (GAL)}.These levels provide a conceptual scaffold for researchers, practitioners, and users to benchmark agent capabilities, track progress over time, and identify the scientific challenges that must be addressed to advance the field.

\begin{figure}
\centering
\includegraphics[width=0.95\textwidth]{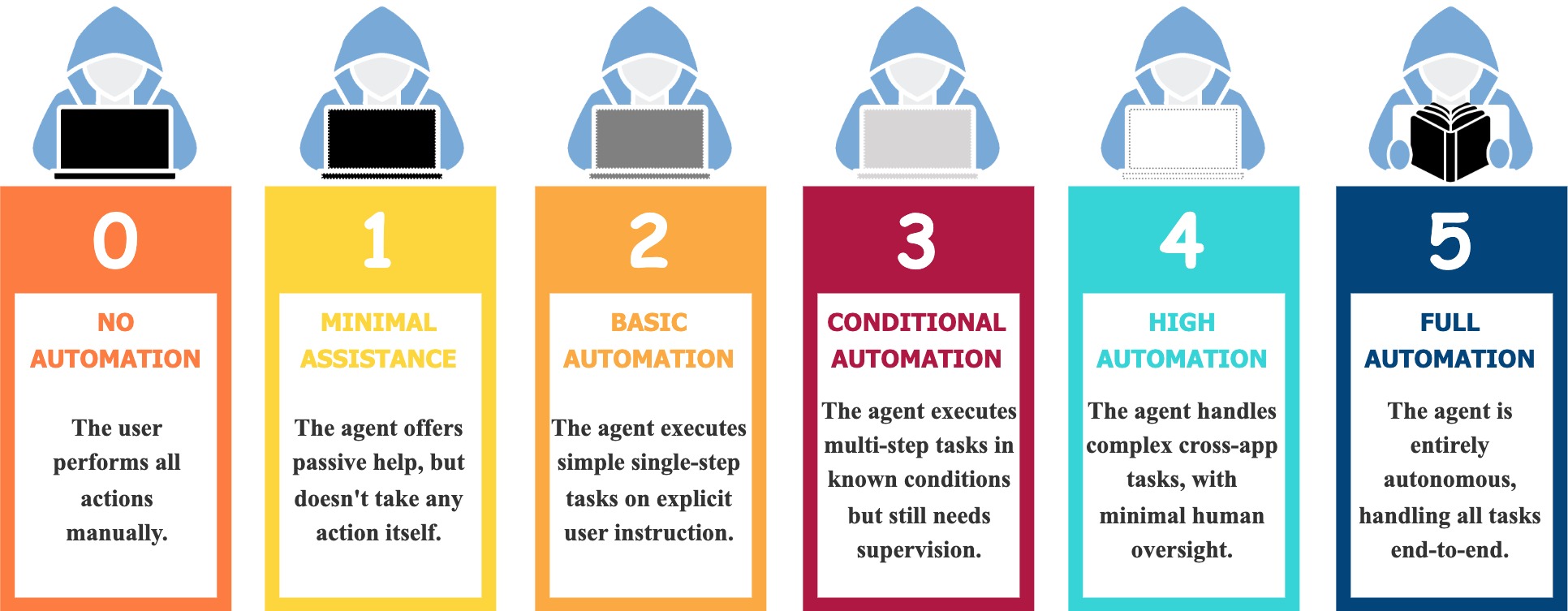}
  \caption{Levels of GUI Agent Autonomy}
  \label{fig:teaser}
\end{figure}

\section{GUI Agent Autonomy Levels}


\subsection{Level 0 (No Automation)}

\textbf{Example}: \textit{``Set up a date event'': The user opens a calendar app, scrolls to a specific date, types in an event, and manually clicks ``Save'', one step at a time. }

At this level, the user performs all actions manually. Every click, scroll, and keyboard input must be initiated by the user, with no automation, assistance, or system memory to support the task. 
The GUI responds only to direct human input, typically through a mouse, keyboard, or touch interface.

\subsection{Level 1 (Minimal Assistance)}
\textbf{Example}: \textit{``Reply to an email'': As the user type, the agent suggests auto-completions based on prior messages, flags cases where an attachment is mentioned but missing, and highlights the ``Send'' button once required fields are complete.}

At Level 1, the agent begins to provide minimal assistance, but it does not execute any actions on the user’s behalf. Instead, it observes the user’s interactions and offers lightweight guidance, such as contextual cues, tooltips, or hints. Basic intelligence emerges at this stage: the agent can analyze GUI elements and provide supportive information, yet it remains strictly advisory rather than operational.

Example tools include smart compose in \textit{Gmail}. 
Such writing assistants analyze user input in real time, suggesting grammar corrections and alternative phrasing within the GUI. They guide the user through underlines and pop-up prompts, but do not make changes unless explicitly approved. 
Similarly, \textit{In-App Hints} and \textit{Onboarding Assistants} provide contextual tooltips and walkthroughs that help users navigate unfamiliar or complex interfaces, but they never interact with the interface directly. 


\subsection{Level 2 (Basic Automation)}
\textbf{Example}: \textit{``Open Instagram'' and ``Click Login button'': The agent executes each command immediately, without clarification or follow-up queries.}

At Level 2, the agent can now perform basic, single-step actions, but only when explicitly instructed by the user. The agent does not make decisions or initiate actions on its own, rather, it listens to user commands and executes them precisely. This marks a transition from passive guidance to active operation. 
Although the user must still decompose tasks into individual steps, the agent can execute these direct operations quickly and reliably, without requiring the user to physically interact with the GUI.

Example tools include \textit{Selenium}~\cite{selenium}, \textit{UI Automator}~\cite{uiautomator}, and \textit{AppleScript}~\cite{applescript_guide}. 
These tools are widely used to automate user interactions in web, mobile, and desktop.
They are script-driven and excel at executing predefined actions in a step-by-step manner (such as opening windows and moving files), but offer limited flexibility when interfaces change or workflows deviate from expected patterns.

\subsection{Level 3 (Conditional Automation)}
\textbf{Example}: \textit{``Set an alarm for 7:30 AM every weekday'': The agent opens the clock application, navigates to the alarm tab, checks whether a recurring alarm already exists, creates one if necessary, and sets the appropriate time and days. If it encounters an unexpected situation, such as requesting additional permissions, it alerts the user and waits for confirmation before proceeding.}

At Level 3, the agent moves beyond responding to isolated commands and begins functioning as a true GUI agent, capable of perceiving interface elements, interpreting user intent, and autonomously executing multi-step workflows.
Such agents can navigate sequences of screens and complete tasks end-to-end. However, their autonomy remains bounded by predefined rules and human-approved flows. That is, the agent behaves like a well-trained assistant: the user specifies the goal, and the agent determines how to achieve it within familiar contexts. It no longer requires step-by-step instruction for every action, yet it is still not ready to operate entirely unsupervised and human oversight remains essential.

For example, \textit{UiPath}~\cite{uipath} and \textit{Automation Anywhere}~\cite{automation_anywhere} are representative Robotic Process Automation (RPA) platforms designed to automate workflows across enterprise applications. They enable agents to replicate user interactions on GUIs and execute multi-step procedures that incorporate conditional logic and basic exception handling. 
In addition, LLM-driven GUI agents like \textit{WebAgent}~\cite{gurreal} and \textit{AppAgent}~\cite{zhang2025appagent} can generate step-by-step plans and ground actions in dynamic interface content.

\subsection{Level 4 (High Automation)}
\textbf{Example}: \textit{``Take the latest sales data from CRM, build a bar chart in Excel, and send it to my boss'': The agent extracts data from the CRM, opens Excel to build the chart, switches to Outlook, writes an email, attaches the file, and sends. If a login prompt appears, such as authentication request in Outlook due to privacy restrictions, the agent asks for user input. Otherwise, the entire workflow is completed without further assistance.}

At Level 4, agents are capable of autonomously handling complex, interconnected tasks across multiple applications, requiring only minimal human oversight. Unlike earlier levels that rely on rigid scripts or closely supervised execution, these agents can understand a task, interpret application context, and adapt to dynamic GUI changes. They make intelligent, context-aware decisions, recover from routine errors, and coordinate multiple subtasks to accomplish higher-level objectives. Although human intervention may still be required in rare situations, such as unexpected edge cases, it is no longer part of the normal workflow.

Some notable agents include \textit{Claude Computer Use Agent}~\cite{claude_computer_use} and \textit{ChatGPT Atlas}~\cite{altas}. These agents can operate within simulated desktop or browser-based environments to carry out multi-step tasks such as web search, information extraction, and cross-page navigation, supported by lightweight state tracking and basic error-recovery mechanisms. 
Similarly, \textit{Manus}~\cite{manus_agent} presents a more advanced autonomous agent that coordinates multiple specialized sub-agents with distinct roles, and executes complex tasks within GUI environments.

\subsection{Level 5 (Full Automation)}
\textbf{Example}: \textit{``I'm onboarding a new employee, please handle everything'': The agent creates new user accounts across systems, completes HR and payroll forms, sets up email and calendar, assigns onboarding documents, schedules orientation meetings, and sends welcome messages.}

This represents the final state: a universal GUI agent. Such an agent would operate across any software, operating system, or interface with minimal prior training. It would be able to interpret underspecified requests, learn from experience, reason over previously unseen GUIs, and optimize task execution in real time. It would be highly adaptable, robust, and capable of proactive behavior. It functions as a true digital collaborator.

\section{Current Status and Future Opportunity}

GUI agents have advanced considerably in recent years. They are now widely deployed in software testing, enterprise workflow automation, and voice-controlled assistive technologies. 
\textbf{Yet most agents in practical use still operate between Level 1 and Level 3}, falling short of genuine autonomy. 
Many agents require extensive configuration, rely on brittle GUI selectors, or depend heavily on back-end APIs rather than directly perceiving, interpreting, and reasoning over visual interfaces.

LLM-driven GUI agents such as Claude Computer Use Agent, ChatGPT Altas, and Manus now demonstrate more flexible, goal-directed behavior in GUI environments, which shows \textbf{a promising step toward Level 4 autonomy}.
However, their ability to generalize across diverse software ecosystems is still constrained, and long-term state and context tracking remain early-stage capabilities. As a result, a human-in-the-loop is still required for oversight, troubleshooting, and correction when workflows deviate from expectations.

\textbf{What, then, comes next?}

Progress toward Level 4, and ultimately Level 5, will require agents that can understand high-level user intentions, operate robustly across heterogeneous interfaces, and adapt in real time. Achieving this will require advances not only in perception and reasoning, but also in several broader challenges, including:

\textbf{Easy to use:}
Today's GUI agents often require detailed configuration or manual scripting. Future agents should be able to generalize across applications and adapt their workflows with minimal human setup. With continued advances in multimodal understanding, agent is likely to become increasingly plug-and-play, capable of operating out of the box across domains such as productivity, finance, and operations.

\textbf{Security:}
As agents will take on more responsibility, security becomes non-negotiable. They must operate within clearly defined permission boundaries, maintain transparent action logs, and be auditable by design. Whether executing internal workflows or interacting with third-party applications, trust will depend on robust policies, sandboxed environments, and fine-grained access control.

\textbf{Privacy:}
Future GUI agents will inevitably handle sensitive user data such as emails, documents, and messages. They must be designed with privacy as a foundational principle, such as minimizing data exposure, supporting on-device execution when feasible, and giving users explicit control over what is stored, shared, or forgotten. Trust in agents will depend on visibility, transparency, and meaningful user consent.

\textbf{Personalization:}
No two users work in the same way.
These agents will be increasingly personalized, such as learning from a user’s preferences, routines, and past actions. Whether it concerns writing styles, file organization habits, or application usage patterns, agents should adapt their behavior to align with individual workflows, providing suggestions and automation that feel tailored. Greater context-awareness, e.g., recognizing when a user is in focus mode, in a meeting, or on the move, will further refine and individualize their support.

\section{Final Conclusion}
GUI agents are undergoing a fundamental transformation. While full autonomy remains out of reach, recent progress, particularly in LLM-powered agents, suggests that the field is beginning to approach Level 4 capabilities, marking a substantial departure from the rule-based systems of the past. 
Nonetheless, achieving Level 5 autonomy, where agents can reliably operate across any software environment without human intervention, remains a distant goal. Reaching this stage will require major advances in perception, reasoning, memory, efficiency, security, and human–agent collaboration. It is an inherently interdisciplinary challenge, including software engineering, natural language processing, human–computer interaction, computer vision, and related fields.



\bibliographystyle{ACM-Reference-Format}
\bibliography{main}

@misc{sae,
  author       = {{SAE International}},
  year         = {2016},
  howpublished          = {\url{https://doi.org/10.4271/J3016_201609}},
}

@inproceedings{gurreal,
  author={Gur, Izzeddin and Furuta, Hiroki and Huang, Austin V and Safdari, Mustafa and Matsuo, Yutaka and Eck, Douglas and Faust, Aleksandra},
  booktitle={The Twelfth International Conference on Learning Representations},
year={2024}
}

@inproceedings{zhang2025appagent,
  author={Zhang, Chi and Yang, Zhao and Liu, Jiaxuan and Li, Yanda and Han, Yucheng and Chen, Xin and Huang, Zebiao and Fu, Bin and Yu, Gang},
  booktitle={Proceedings of the 2025 CHI Conference on Human Factors in Computing Systems},
  pages={1--20},
  year={2025}
}

@misc{claude_computer_use,
  author       = {{Anthropic}},
  year         = {2024},
  howpublished          = {\url{https://www.anthropic.com/news/3-5-models-and-computer-use}},
}

@misc{altas,
  author       = {{OpenAI}},
  year         = {2025},
  howpublished          = {\url{https://openai.com/index/introducing-chatgpt-atlas/}},
}

@misc{manus_agent,
  author       = {{Monica}},
  year         = {2025},
  howpublished          = {\url{https://manus.im/}},
}

@misc{uipath,
  author       = {{UiPath}},
  year         = {2005},
  howpublished          = {\url{https://www.uipath.com}},
}

@misc{automation_anywhere,
  author       = {{Automation Anywhere}},
  year         = {2003},
  howpublished = {\url{https://www.automationanywhere.com}},
}

@misc{applescript_guide,
  author       = {{Apple}},
  year         = {2016},
  howpublished          = {\url{https://developer.apple.com/library/archive/documentation/AppleScript/Conceptual/AppleScriptLangGuide}},
}

@misc{selenium,
  author       = {{Selenium}},
  year         = {2004},
  howpublished          = {\url{https://www.selenium.dev}},
}

@misc{uiautomator,
  author       = {{Google}},
  year         = {2026},
  howpublished          = {\url{https://developer.android.com/training/testing/ui-automator}},
}

@misc{comet,
  author       = {{Perplexity}},
  year         = {2025},
  howpublished          = {\url{https://www.perplexity.ai/comet/}},
}

@article{qin2025ui,
  author={Qin, Yujia and Ye, Yining and Fang, Junjie and Wang, Haoming and Liang, Shihao and Tian, Shizuo and Zhang, Junda and Li, Jiahao and Li, Yunxin and Huang, Shijue and others},
  journal={Preprint at \url{https://arxiv.org/abs/2501.12326}},
  year={2025}
}

@inproceedings{rawlesandroidworld,
  author={Rawles, Christopher and Clinckemaillie, Sarah and Chang, Yifan and Waltz, Jonathan and Lau, Gabrielle and Fair, Marybeth and Li, Alice and Bishop, William E and Li, Wei and Campbell-Ajala, Folawiyo and others},
  booktitle={The Thirteenth International Conference on Learning Representations}
}

@article{xie2024osworld,
  author={Xie, Tianbao and Zhang, Danyang and Chen, Jixuan and Li, Xiaochuan and Zhao, Siheng and Cao, Ruisheng and Hua, Toh J and Cheng, Zhoujun and Shin, Dongchan and Lei, Fangyu and others},
  journal={Advances in Neural Information Processing Systems},
  volume={37},
  pages={52040--52094},
  year={2024}
}

\end{document}